%% file: 0Paper.tex
\documentclass[sigconf]{acmart}

\AtBeginDocument{%
  }

\usepackage{marvosym}
\usepackage[labelformat=simple]{subcaption}

\usepackage{enumitem}
\usepackage[skip=0.5\baselineskip]{caption}
\usepackage{bm}
\usepackage{algorithmicx,algorithm}
\usepackage{algpseudocode}
\usepackage{arydshln}
\usepackage{float}
\usepackage{overpic}
\usepackage{float}  
\usepackage{subfloat}
\usepackage{graphicx} 			
\usepackage{subcaption}			
\usepackage{bm}
\usepackage{multirow}
\usepackage{graphicx}
\usepackage{subcaption}
\usepackage{colortbl}
\usepackage{enumitem}
\usepackage{stfloats} 
\usepackage{url}
\usepackage[normalem]{ulem}
\useunder{\uline}{\ul}{}
\usepackage[skip=0.5\baselineskip]{caption}
\definecolor{editred}{RGB}{150,84,84}

\copyrightyear{2024}
\acmYear{2024}
\setcopyright{acmlicensed}
\acmConference[CIKM '24] {Proceedings of the 33rd ACM International Conference on Information and Knowledge Management}{October 21--25, 2024}{Boise, ID, USA.}
\acmBooktitle{Proceedings of the 33rd ACM International Conference on Information and Knowledge Management (CIKM '24), October 21--25, 2024, Boise, ID, USA}
\acmISBN{979-8-4007-0436-9/24/10}
\acmDOI{10.1145/3627673.3679842}
\usepackage[medium,compact]{titlesec}
\usepackage{xspace}

\newcommand{\eg}{\emph{e.g.,}\xspace}
\newcommand{\ie}{\emph{i.e.,}\xspace}

\newcommand{\name}{ROK\xspace}
\newcommand{\firstcategory}{feature interaction-based methods\xspace}
\newcommand{\secondcategory}{user behavior modeling methods\xspace}
\newcommand{\thirdcategory}{item-level retrieval-based methods\xspace}
\newcommand{\fourthcategory}{sample-level retrieval-based methods\xspace}

\author{Huanshuo Liu}
\authornote{This work is done during the internship at Huawei Noah's Ark Lab.}
\affiliation{%
  \institution{National University of Singapore}
  \city{Singapore}
  \country{Singapore}
}
\email{liuhsh35@mail2.sysu.edu.cn}

\author{Bo Chen}
\affiliation{%
  \institution{Huawei Noah’s Ark Lab}
  \city{Shenzhen}
  \country{China}
}
\email{chenbo116@huawei.com}

\author{Menghui Zhu}
\affiliation{%
  \institution{Huawei Noah’s Ark Lab}
  \city{Shenzhen}
  \country{China}
}
\email{zhumenghui1@huawei.com}

\author{Jianghao Lin}
\affiliation{%
  \institution{Shanghai Jiao Tong University}
  \city{Shanghai}
  \country{China}
}
\email{chiangel@sjtu.edu.cn}

\author{Jiarui Qin}
\affiliation{%
  \institution{Huawei Noah’s Ark Lab}
  \city{Shenzhen}
  \country{China}
}
\email{qinjr@apex.sjtu.edu.cn}

\author{Hao Zhang}
\affiliation{%
  \institution{Huawei Noah’s Ark Lab}
  \city{Singapore}
  \country{Singapore}
}
\email{hzhang26@outlook.com}

\author{Yang Yang}
\affiliation{%
  \institution{Huawei Noah’s Ark Lab}
  \city{Shanghai}
  \country{China}
}
\email{yangyang590@huawei.com}

\author{Ruiming Tang}
\authornote{Ruiming Tang is the corresponding author.}
\affiliation{%
  \institution{Huawei Noah’s Ark Lab}
  \city{Shenzhen}
  \country{China}
}
\email{tangruiming@huawei.com}

\renewcommand{\shortauthors}{Huanshuo Liu et al.}

\begin{document}

\title{Retrieval-Oriented Knowledge for Click-Through Rate Prediction}
\begin{abstract}
Click-through rate (CTR) prediction is crucial for personalized online services. Sample-level retrieval-based models, such as RIM, have demonstrated remarkable performance. However, they face challenges including inference inefficiency and high resource consumption due to the retrieval process, which hinder their practical application in industrial settings. To address this, we propose a universal plug-and-play \underline{r}etrieval-\underline{o}riented \underline{k}nowledge (\textbf{\name}) framework that bypasses the real retrieval process. The framework features a knowledge base that preserves and imitates the retrieved \& aggregated representations using a decomposition-reconstruction paradigm. Knowledge distillation and contrastive learning optimize the knowledge base, enabling the integration of retrieval-enhanced representations with various CTR models.
Experiments on three large-scale datasets demonstrate \name's exceptional compatibility and performance, with the neural knowledge base serving as an effective surrogate for the retrieval pool. \name surpasses the teacher model while maintaining superior inference efficiency and demonstrates the feasibility of distilling knowledge from non-parametric methods using a parametric approach. These results highlight \name's strong potential for real-world applications and its ability to transform retrieval-based methods into practical solutions.
Our implementation code is available to support reproducibility\footnote{\url{https://github.com/HSLiu-Initial/ROK.git}}.
\vspace{-1.5mm}
\end{abstract}



\maketitle
\input{1Introduction}

\input{2ProposedMethod}
\input{3Experiments}
\input{4Deployment}
\input{5Conslusion}
\bibliographystyle{ACM-Reference-Format}
\bibliography{reference}


\end{document}

%% file: 1Introduction.tex
\vspace{-10pt}

\section{Introduction}

Click-through rate (CTR) prediction is a key component of many personalized online services, such as recommender systems~\cite{xi2023bird,wang2023plate} and web search~\cite{lin2021graph,fu2023f,dai2021adversarial,liu2024ctrla}. 
It aims to estimate the probability of a user's click given a particular context~\cite{zhang2021deep}. 
The CTR models can be mainly classified into two categories.
The first category is the \firstcategory. 
The core idea of these methods is to capture the high-order feature interactions across multiple fields with different operators (\eg product~\cite{qu2018product,DCNv2,guo2017deepfm,edcn}, convolution~\cite{cfm,liu2019feature}, and attention~\cite{song2019autoint,xiao2017attentional}).
The second category is the \secondcategory that leverage different architectures (\eg RNN~\cite{hidasi2018recurrent,hidasi2015session}, CNN~\cite{tang2018personalized}, attention~\cite{zhou2019deep,zhou2018deep}, memory bank~\cite{pi2019practice,ren2019lifelong}) to extract informative knowledge from user behavior sequences for final CTR prediction.

Due to the problem of distribution shift and the daily generation of vast amounts of interaction data in industrial applications, using all the data for training becomes impractical. This results in a significant waste of historical data, as traditional methods cannot effectively utilize it.
To address this issue, retrieval-based methods have been proposed. UBR4CTR~\cite{qin2020user} and SIM~\cite{pi2020search} retrieve useful behaviors from the user's behavior history (\ie clicked items), reducing the potential noise in user behavior sequences.
Subsequent studies~\cite{li2022adversarial} improve these methods' efficiency using hashing functions~\cite{chen2021end} and parallel retrieval execution~\cite{cao2022sampling} during inference.
Recent studies~\cite{qin2021retrieval,du2022learning} further expand these retrieval-based methods from \textbf{item-level} retrieval to \textbf{sample-level} retrieval, making them applicable to general CTR prediction settings.
Instead of retrieving similar items from the \textbf{user history}, RIM~\cite{qin2021retrieval} adopts the idea of $k$ nearest neighbor ($k$NN) and designs a sample-centric retrieval method, which aggregates the relevant data samples retrieved from the search pool (\eg the whole training dataset).

\begin{figure}[htbp]   
    \centering
\includegraphics[width=\linewidth, height=\textheight, keepaspectratio]{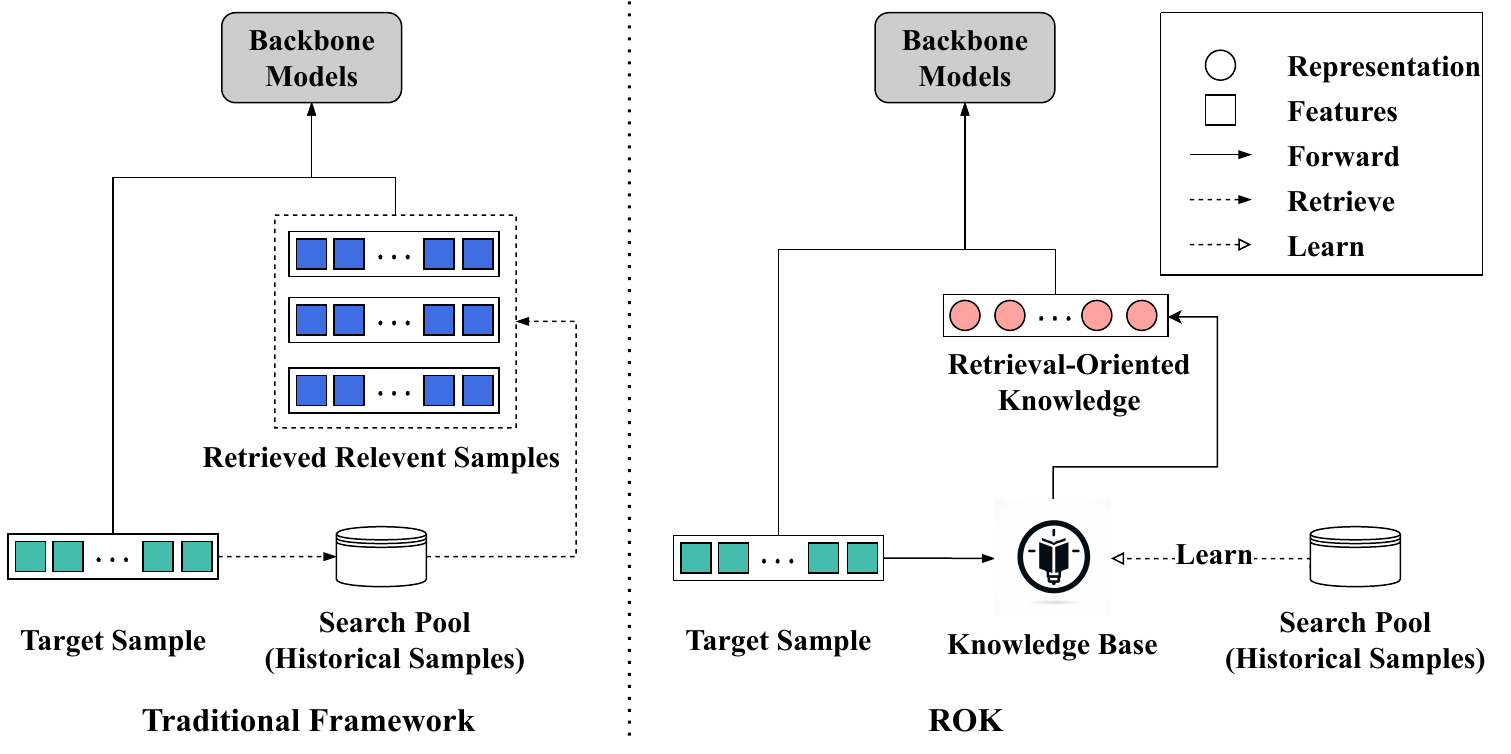}
    \caption{Comparison between traditional sample-level retrieval framework and \name.}
    \label{fig: Traditional}
\end{figure}

Although \fourthcategory brings impressive performance enhancement, they have to perform instance-wise comparisons between the target data sample and candidate samples in the search pool (usually at the million or even billion level), leading to \textbf{extreme inefficiency problems during inference}, as illustrated in the left part of Figure~\ref{fig: Traditional}. While follow-up studies~\cite{zheng2023dense,Li_2024} have attempted to enhance retrieval mechanisms through vector retrieval techniques, they have instead created further challenges and failed to meet the necessary inference speed for deployment.
Moreover, \fourthcategory faces several other significant challenges. These include \textbf{high resource consumption} due to the linear increase in retrievers with request volume, as well as a \textbf{divergent inference process} that requires an inference-retrieval-inference cycle. This necessitates significant optimizations, which makes it challenging for industrial applications.

This work introduces the \underline{R}etrieval-\underline{O}riented \underline{K}nowledge (\name) framework to address the problems of sample-level retrieval-based methods by bypassing the real retrieval process. The framework operates in two stages: Retrieval-Oriented Knowledge Construction and Knowledge Utilization.
In the first stage, we pre-train a sample-level retrieval-based method, such as RIM~\cite{qin2021retrieval}. 
To bypass the real retrieval process, \name learns a neural network-based Knowledge Base to store the retrieval-oriented knowledge from this pre-trained method. Specifically, \name imitates the aggregated representations from the pre-trained retrieval-based method via a Retrieval Imitation module, as illustrated in Figure~\ref{fig: Traditional}. The Knowledge Base, built on a novel \textbf{decomposition-reconstruction paradigm}, where a Retrieval-Oriented Embedding Layer captures the \textit{feature-wise} embedding and a Knowledge Encoder reconstructs the \textit{instance-wise} aggregated representations (\ie retrieval-enhanced representation). 
In this way, instead of the time-consuming retrieval, efficient inference of neural networks can be adopted to get the approximated aggregated representations. 
Additionally, we introduce a Contrastive Regularization module to ensure learning stability and prevent model collapse.
At the Knowledge Utilization stage, \name designs two approaches (i.e., instance-wise and feature-wise) to integrate the retrieval-enhanced representations with various CTR models, thus providing high inference efficiency.

The main contributions of this paper are as follows:
\begin{list}{\labelitemi}{\leftmargin=1em}
    \setlength{\topmargin}{0pt}
    \setlength{\itemsep}{0em}
    \setlength{\parskip}{0pt}
    \setlength{\parsep}{0pt}

\item \name is a novel framework that leverages knowledge distillation to efficiently utilize large amounts of historical data and retrieval-oriented knowledge for training and lightweight deployment. To the best of our knowledge, \name is the first such framework. It transforms \fourthcategory into a practical solution by using a neural network-based Knowledge Base to store retrieval-enhanced representations, eliminating the need for time-intensive retrieval during inference. This demonstrates the feasibility of distilling knowledge from non-parametric models and learning their inherent knowledge using a parametric approach.
\item \name optimizes the Knowledge Base using a combination of knowledge distillation and contrastive learning. This approach allows for the seamless integration of retrieval-enhanced representations with various CTR models at both the instance and feature levels, enhancing their performance and compatibility.
\item Extensive experiments across three large-scale datasets demonstrate \name's exceptional compatibility and performance. The results show that \name significantly enhances the performance of various CTR methods, indicating that the neural Knowledge Base serves as an effective and compact surrogate for the search pool. Although the knowledge distillation foundation initially suggests that \name may only approach the teacher model's performance, the introduction of contrastive regularization enables \name to surpass the sample-level retrieval-based teacher model. 
This highlights \name's strong potential for real-world applications.
\end{list}

%% file: 2ProposedMethod.tex
\begin{figure*}[htbp]   
    \centering
\includegraphics[width=\linewidth, height=\textheight, keepaspectratio]{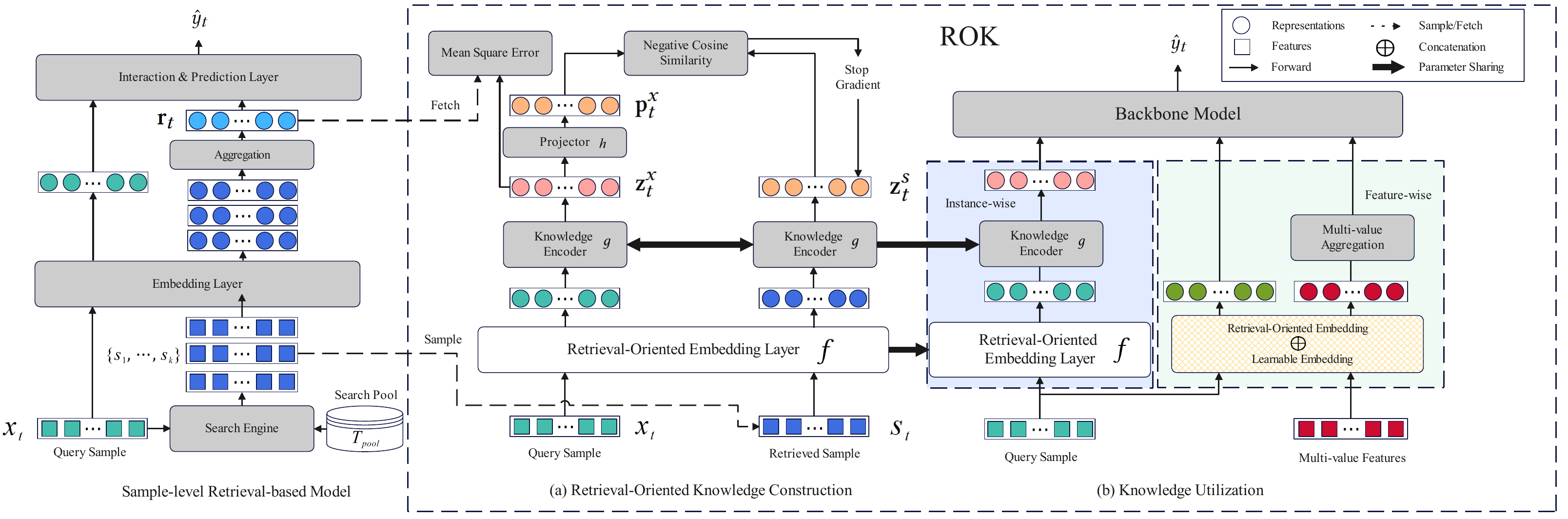}
    \caption{Overall framework of \name. The process begins with the Retrieval-Oriented Knowledge Construction stage, where a sample-level retrieval-based model has already been pre-trained, and the retrieval and aggregation results are offline pre-stored. A tailored knowledge base is then designed to imitate the aggregation results, with Contrastive Regularization ensuring stabilization during the training phase. This design allows the time-consuming retrieval \& aggregration process to be substituted with the swift forward propagation of a neural network. In the subsequent Knowledge Utilization stage, the framework introduces the retrieval-enhanced representation into backbone CTR models in both an instance-wise and feature-wise approach, ensuring the maintenance of sample-level retrieval abilities while circumventing the complexities of retrieval \& aggregation operations.}
    \label{fig: Overall Framework}
\end{figure*}
\vspace{-10pt}
\section{Related Work}
\subsection{CTR Prediction}
For click-through rate (CTR) prediction, models can essentially be divided into two main categories: those focusing on feature interaction and those centering on user behavior modeling.
The first category is the \firstcategory, evolving from foundational works such as POLY2~\cite{chang2010training} and Factorization Machines (FM)~\cite{rendle2010factorization}. With the integration of Deep Neural Networks (DNNs), a variety of sophisticated deep feature interaction models have been proposed. These models aim to capture high-order feature interactions across different fields by employing various operations, such as the product operation~\cite{qu2018product,DCNv2,guo2017deepfm,edcn,inttower}, convolution~\cite{cfm,liu2019feature}, and attention mechanisms~\cite{song2019autoint,xiao2017attentional}. The key innovation of these models lies in their ability to identify and utilize complex interactions between a multitude of features to enhance the predictive performance of CTR models.

Furthermore, user behavior modeling is another core technique for CTR prediction that mines user preferences from historical interaction behaviors meticulously. To better extract informative knowledge from a user’s behavior sequence, various network structures have been utilized, including Recurrent Neural Networks (RNNs), Convolutional Neural Networks (CNNs), Attention Networks, and Memory Networks. GRU4Rec~\cite{hidasi2015session} designs Gated Recurrent Units (GRUs) to capture the preference-evolving relationship, while Caser~\cite{tang2018personalized} leverages the horizontal and vertical convolution to model skip behaviors at both the union-level and point-level. Moreover, the attention mechanism is the most popular method for modeling item dependencies, and several influential works have been proposed, including SASRec~\cite{kang2018self}, DIN~\cite{zhou2018deep}, DIEN~\cite{zhou2019deep}, and BERT4Rec~\cite{sun2019bert4rec}. Among these, DIN and DIEN leverage the target-attention network to identify important historical items, while SASRec and BERT4Rec apply the self-attention network with Transformer architecture to excavate behavior dependencies. Additionally, memory-based methods~\cite{jiang2020multiplex} are also proposed to store user behavior representations in a read-write manner. 

\subsection{Retrieval-Augmented Recommendation}
To further enhance the performance of CTR prediction, retrieval-augmented recommendation is proposed, where the most relevant information is retrieved from historical data. Specifically, UBR4CTR~\cite{qin2020user} and SIM~\cite{ren2019lifelong} are designed to retrieve beneficial behaviors from the user's extremely long historical behavior sequence. UBR4CTR deploys the search engine method, while SIM uses hard search and soft search approaches. To make the search procedure end-to-end, ETA~\cite{chen2022efficient} is proposed by leveraging the SimHash algorithm to map user behavior into a low-dimensional space, hence achieving learnable retrieval. Moreover, recent works further extend retrieval-augmented recommendation from \textbf{item-level} retrieval to \textbf{sample-level} retrieval by retrieving informative samples. RIM~\cite{qin2021retrieval} is the first to deploy this method, leveraging the search engine to retrieve several relevant samples from the search pool and perform neighbor aggregation. PET~\cite{du2022learning} and DERT~\cite{zheng2023dense} have respectively made improvements in the interaction and retrieval mechanisms of neighboring samples. PET constructs a hypergraph over the retrieved data samples and performs message propagation to improve the target data representations for final CTR prediction. DERT utilizes vector retrieval to speed up the retrieval process.

\section{Preliminary}

\subsection{CTR Prediction}

In CTR prediction, each data sample is denoted as $s_t=(x_t,y_t)$, where$\ x_t=\{c_i^t\}_{i=1}^F$, $F$ is the number of discrete features\footnote{Continuous features are usually discretized with various methods~\cite{autodis}.}, and $y_t$ is the label. Thus a dataset with $N$ samples can be expressed as $T=\left\{s_{t}\right\}_{t=1}^{N}$. 
The goal of the CTR prediction is to estimate the click probability of a specific sample: $\hat{y}_{t}=G\left(x_{t} ; \mathbf{\theta}\right)$, where $G$ is the CTR model with learnable parameters $\theta$.

Apart from the traditional training/validation/testing splits for CTR prediction, retrieval-based methods further require a search pool $T_{pool}$, which might overlap with the training set $T_{train}$ according to different settings.
The search pool $T_{pool}$ is constructed to provide useful knowledge for downstream CTR prediction towards the target sample $x_t$, which can be formulated as:
\begin{equation}
    \hat{y}_{t}=G\left(x_{t}, R\left(x_{t}\right) ; \mathbf{\theta}\right),
\end{equation} 
where $R\left(x_{t}\right)$ is the retrieved knowledge of $x_t$.
For \emph{\textbf{item-level}} retrieval~\cite{qin2020user,pi2020search}, the knowledge $R\left(x_{t}\right)$ is the retrieved $k$ user behaviors.
For \emph{\textbf{sample-level}} retrieval~\cite{qin2021retrieval,du2022learning}, the knowledge $R\left(x_{t}\right)$ is the retrieved $k$ nearest data samples.
As for our proposed \emph{\textbf{\name}}, the knowledge $R\left(x_{t}\right)$ is the learned retrieval-enhanced representation.

After obtaining the click prediction $\hat{y}_{t}$, the parameters $\theta$ are optimized by minimizing the binary cross-entropy (BCE) loss:
\begin{equation}
\label{eq:modeling}
    \mathcal{L}_t = y_t\log\hat{y}_t + (1-y_t)\log(1-\hat{y}_t).
\end{equation}

\subsection{Sample-level Retrieval-based Methods}

For illustration purposes, we abstract the sample-level retrieval approach in the left part of Figure \ref{fig: Overall Framework}.
The target sample $x_t$ will be considered as a query to retrieve the top-K neighboring samples $\{s_1,\cdots, s_K\}$ from the search pool $T_{pool}$ through the search engine. After retrieval, an aggregation layer is used to aggregate the features and labels of the retrieved samples, resulting in a dense knowledge representation $\mathbf{r}_t$. 
For example, RIM~\cite{qin2021retrieval} deploys an attentive aggregation layer, while PET~\cite{du2022learning} constructs a hypergraph over the retrieved samples, and performs message propagation to aggregate the representations. 
Finally, the aggregated representation $\mathbf{r}_t$ is sent to the prediction module together with the target sample representation. The prediction module usually contains components for feature interaction modeling~\cite{edcn,guo2017deepfm,song2019autoint} or user behavior modeling~\cite{zhou2019deep,zhou2018deep}.
\section{Methodology}
\subsection{Overview of \name}
Despite their superior performance, sample-level retrieval-based methods suffer from challenges, including inference inefficiency and high resource consumption, due to the time-consuming online retrieval process.
To this end, we propose a novel Retrieval-Oriented Knowledge (\name) framework, where a knowledge base is built to imitate the aggregated retrieval knowledge $\mathbf{r}_t$.
As shown in Figure~\ref{fig: Overall Framework}, ROK consists of two stages: (1) Retrieval-Oriented Knowledge Construction, and (2) Knowledge Utilization.

In the first stage of ROK--\emph{retrieval-oriented knowledge construction}--we pre-train a sample-level retrieval-based model\footnote{In this work, as our research primarily focuses on inference efficiency, we choose a simple yet effective option that provides sufficiently good results: RIM~\cite{qin2021retrieval}, which is one of the foundational works in \fourthcategory. The pre-training process strictly follows the original work~\cite{qin2021retrieval}. It is important to note that our proposed method can be generalized to all \fourthcategory methods, extending its applicability and potential impact.}, followed by designing a delicate \emph{knowledge base} with a retrieval-oriented embedding layer and a knowledge encoder.
This is designed to imitate the retrieval \& aggregation results (\ie $\mathbf{r}_t$) of the pre-trained model by directly generating the final aggregated representation $\mathbf{z}_t^x$ through the knowledge encoder.
In this way, when online serving, the time-consuming retrieval process could be replaced by a simpler and faster forward propagation of a neural network (\ie knowledge base).
We adopt the mean square error loss for knowledge imitation and design a contrastive regularization loss to stabilize the learning process and prevent collapse~\cite{jing2021understanding,shen2022connect}.

In the \emph{knowledge utilization} stage, we propose to integrate the plug-and-play retrieval-enhanced representations into arbitrary backbone CTR models in both \emph{instance-wise} and \emph{feature-wise} manner.
Hence, we retain the sample-level retrieval capacity, while avoiding the online overhead brought by retrieval \& aggregation operations.

Next, we will first elaborate on the network structure design of the \textbf{knowledge base} for knowledge imitation. Then, we give the detailed training strategies for the two stages (\ie retrieval-oriented knowledge construction and knowledge utilization).

\subsection{Structure Design of Knowledge Base}

\begin{figure}[htbp]   
    \centering
\includegraphics[width=\linewidth, height=\textheight, keepaspectratio]{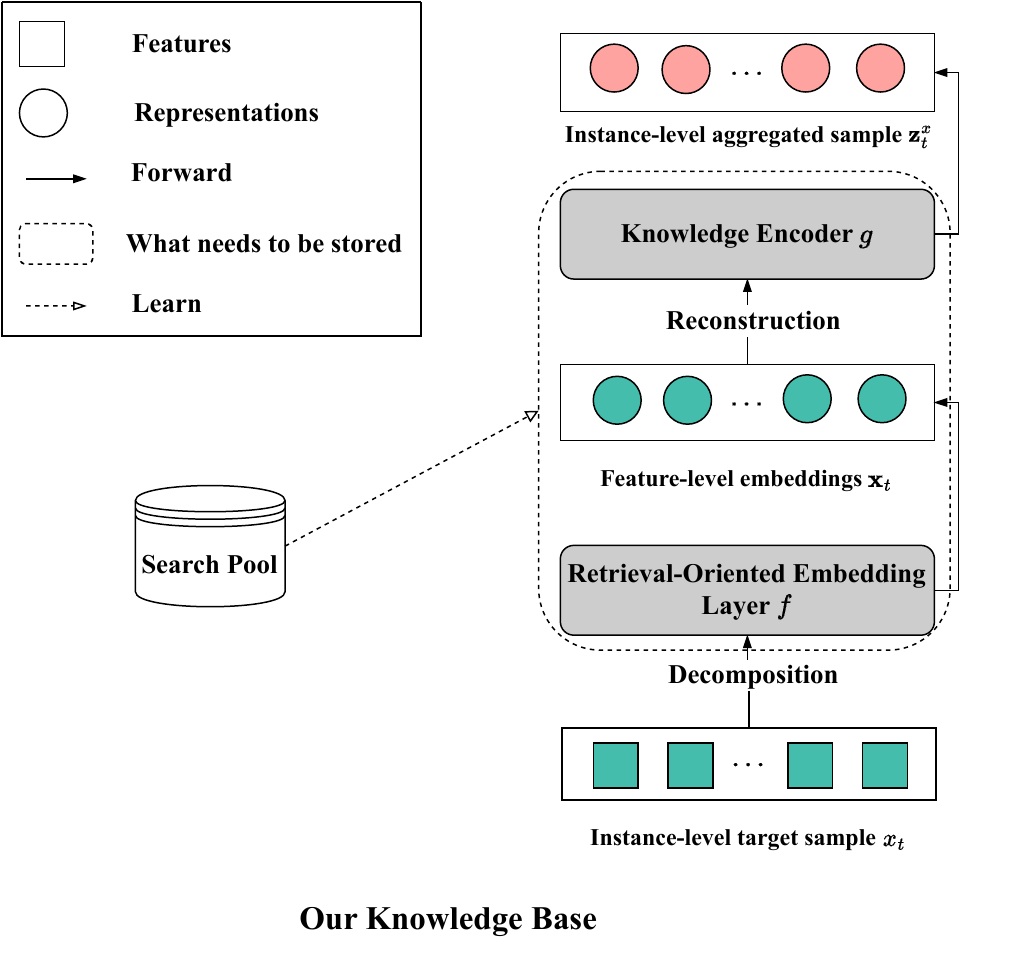}
    \caption{Structure Design of Knowledge Base.}
    \label{fig: Knowledge Base}
\end{figure}

For \fourthcategory~\cite{qin2021retrieval,du2022learning}, the overhead of inference inefficiency and high resource consumption is mainly caused by the retrieval \& aggregation operations, which are uncacheable and unavoidable during the online inference if a brand-new data sample comes. 
To this end, we manage to distill knowledge from non-parametric retrieval process using a parametric approach.
To be specific, we propose a novel \textbf{decomposition-reconstruction paradigm}, where the aggregated representation $\mathbf{r}_t$ is first decomposed into feature-level embeddings, and then reconstructed via a neural encoder.
As illustrated in ~\ref{fig: Knowledge Base}, our proposed knowledge base comprises two key modules: (1) retrieval-oriented embedding layer $f$, and (2) knowledge encoder $g$.

Initially, we decompose the query $x_t$ into \emph{feature-wise} embeddings via the retrieval-oriented embedding layer $f$.
Then, the obtained feature-wise embeddings are fed into the knowledge encoder $g$ to reconstruct the \emph{instance-wise} aggregated representation $\mathbf{z}_t^x=g(f(x_t))$. Specifically, a learnable embedding layer functions as the retrieval-oriented embedding layer. The architecture of knowledge encoder $g$ can be chosen arbitrarily (\eg transformer~\cite{vaswani2017attention}).
For simplicity, we adopt multi-layer perceptron (MLP) in this paper. This approach reduces the time complexity of sample-level retrieval to $\mathcal{O}(1)$.
Since we introduce the retrieval-oriented embedding layer $f$, we will cut down the embedding size of $f$ from $d$ to $d/2$ for fair comparison in space complexity.
With the help of the learned knowledge base, online retrieval \& aggregation operations can be avoided.

\subsection{Retrieval-Oriented Knowledge Construction}

To construct the knowledge base, we introduce two modules: Retrieval Imitation and Contrastive Regularization.


\subsubsection{Retrieval Imitation}
\label{sec:retrieval imitation}

After we obtain the reconstructed representation $\mathbf{z}_t^x$ from the knowledge base for query sample $x_t$. We use mean square error (MSE) loss to train the knowledge base, enabling it to imitate the aggregated representation $\mathbf{r}_t$:
\begin{equation}
\mathcal{L}_{imit}=\mbox{MSE}\left(\mathbf{z}^x_t,\mathbf{r}_t\right).
\end{equation}
This knowledge imitation approach acts as a knowledge distillation process, where we distill knowledge from a non-parametric retrieval process using a parametric approach. By extracting and injecting the retrieval-oriented knowledge from the pre-trained model into the learnable knowledge base, we enable implicit sample-level retrieval capability.



\subsubsection{Contrastive Regularization}
\label{sec:Contrastive Regularization}

Besides, we have intricately integrated a contrastive regularization loss to ensure stability in the learning process and avert potential collapse~\cite{jing2021understanding,shen2022connect}. Notably, \fourthcategory~\cite{qin2021retrieval,du2022learning} ingeniously produces positive samples during the retrieval process. This can be perceived as a unique data augmentation technique, offering a localized positive perspective for contrastive learning. Consequently, \name employs the SimSiam~\cite{chen2021exploring} framework, utilizing a free negative samples scheme for regularization. While the traditional \fourthcategory predominantly harness global features from the aggregated neighboring samples, our use of contrastive regularization allows for a more detailed extraction of local features from these samples.


As shown in Figure~\ref{fig: Overall Framework}(a), the most correlated neighboring sample $s_t$ is selected from the $K$ retrieved samples $\{s_1,\cdots, s_K\}$ by the \fourthcategory.
Then, the target sample $x_t$ and selected neighboring sample $s_t$ are fed into the knowledge base to obtain the reconstructed representations $\mathbf{z}^x_t$ and $\mathbf{z}^s_t$, respectively.
To prevent collapse in the absence of negative samples~\cite{zhang2022does}, a projector $h$ is employed to generate the projected representation $\mathbf{p}^x_t$, $\mathbf{p}^s_t$ from $\mathbf{z}^x_t$, $\mathbf{z}^s_t$.
\begin{align}
     \mathbf{p}^x_t &\triangleq h\left(\mathbf{z}^x_t\right) \triangleq h\left(g\left(f\left(x_{t}\right)\right)\right),\\
     \mathbf{p}^s_t &\triangleq h\left(\mathbf{z}^s_t\right) \triangleq h\left(g\left(f\left(s_{t}\right)\right)\right).
     \vspace*{-0.5\baselineskip}  
\end{align}
The cosine similarity between projected representation $\mathbf{p}^x_t$ and reconstructed representations $\mathbf{z}^s_t$ is defined as:
\begin{equation}
    \mathcal{D}\left(\mathbf{p}^x_t, \mathbf{z}^s_t\right)=\frac{\mathbf{p}^x_t}{\left\|\mathbf{p}^x_t\right\|_{2}} \cdot \frac{\mathbf{z}^s_t}{\left\|\mathbf{z}^s_t\right\|_{2}},
\end{equation}
where $\left\|\cdot\right\|_{2}$ is the $l_2$ norm. Besides, the gradient of $\mathbf{z}^s_{t}$ is stopped when computing the loss~\cite{zhang2022does,chen2021exploring}.
To avoid spatial bias, motivated by the Jensen–Shannon (JS) divergence, a symmetric loss is adopted and the final contrastive regularization loss is:
\begin{equation}
    \mathcal{L}_{contra}=-\left(\frac{1}{2} \mathcal{D}\left(\mathbf{p}^x_t, \mathbf{z}^s_t\right)+\frac{1}{2} \mathcal{D}\left(\mathbf{p}^s_t, \mathbf{z}^x_t\right)\right).
\end{equation}

Hence, The overall objective for \name in the \emph{retrieval-oriented knowledge construction} stage is:
\begin{equation}
    \mathcal{L}=(1-\alpha)\cdot\mathcal{L}_{imit}+\alpha\cdot\mathcal{L}_{contra},
\end{equation}
where $\alpha$ adjusts the ratio of the two loss terms.
After the retrieval-oriented knowledge construction stage, the knowledge base will be retained for the follow-up knowledge utilization stage.

\subsection{Knowledge Utilization}
\label{sec: knowledge utilization}

In the knowledge utilization stage, the plug-and-play retrieval-enhanced representations from the learned knowledge base can be integrated into various backbone CTR models. 
When integrated with backbone models, all parameters of \name are frozen—a strategy that outperforms others, as detailed in Section~\ref{sec: update_strategy}.
As illustrated in Figure~\ref{fig: Overall Framework}(b), we present two approaches for knowledge utilization, which are mutually compatible to boost the final CTR prediction:
\begin{itemize}[leftmargin=10pt]
    \item \textbf{Feature-wise retrieval-enhanced} method leverages the retrieval-oriented embedding layer $f$ to obtain the retrieval-enhanced feature-wise embeddings for the target sample. Specifically, this method integrates the retrieval-oriented embedding layer $f$ with the original learnable feature embedding of the backbone model to form the final embedding layer\footnote{
The embedding layer accounts for a large proportion of the total number of parameters in CTR models. For a fair comparison in model parameters, the embedding size of both the retrieval-oriented embedding layer and the original learnable feature embedding of the backbone model is set to $d/2$ compared with other models.} \label{sec:half}.
    \item \textbf{Instance-wise retrieval-enhanced} method employs the knowledge base to generate a retrieval-enhanced instance-level aggregated representation $\mathbf{z}_t^x$ for each sample. Typically, various backbone models produce their unique representations—for example, RIM's aggregated representations and DIEN's interest states. These unique representations are then concatenated with the feature embeddings and the combined results are fed into the interaction and MLP (Multi-Layer Perceptron) layers.
    \begin{align}
\label{eq:knowledge utilization}
\left\{
\begin{array}{ll}
\mathbf{z}_t^x &= g(f(x_t)) \\
\hat{y}_t &= \text{backbone} (\phi(x_t) \oplus \mathbf{z}_t^x)
\end{array},
\right.
\end{align} where $\phi$ is utilized to extract the unique representations of the backbone model.
    
\end{itemize}


%% file: 3Experiments.tex
\section{Experiments}
In this section, we present the experimental settings and corresponding results in detail. The experiments are conducted on three large-scale datasets, including Tmall, Taobao, and Alipay. 
To gain more insights into ROK, we endeavor to address the following research questions (RQs) in this section.

\begin{list}{\labelitemi}{\leftmargin=1em}
    \setlength{\topmargin}{0pt}
    \setlength{\itemsep}{0em}
    \setlength{\parskip}{0pt}
    \setlength{\parsep}{0pt}
\item \textbf{RQ1:} How does \name's performance compare to that of traditional CTR models and the teacher model?
\item \textbf{RQ2:} How compatible is ROK with other backbone models?
\item \textbf{RQ3:} How does the knowledge captured by ROK contribute to improving performance?
\item \textbf{RQ4:} How do the chosen update strategy and positive sample selecting strategy affect ROK's performance?
\item \textbf{RQ5:} How does the specified hyperparameter influence ROK's performance?
\end{list}

\subsection{Experimental Settings}
\subsubsection{Datasets.}
The evaluations are conducted on three widely-recognized public datasets: Tmall, Taobao, and Alipay. Comprehensive statistics for these datasets are presented in Table \ref{tab:dataset statistics}. To be specific, we tally the counts of users, items, samples, fields, and categories for each dataset. It's worth noting that the number of features refers to the count of unique feature values.

\begin{table}[htbp]
\caption{The dataset statistics.}
\label{tab:dataset statistics}
\centering
\begin{tabular}{@{}llllll@{}}
\toprule
Dataset & Users \#  & Items \#  & Samples \#  & Fields \#  & Features \# \\
\midrule
Tmall   & 424,170  & 1,090,390 & 54,925,331 & 9       & 1,529,676    \\
Taobao  & 987,994  & 4,162,024 & 100,150,807 & 4       & 5,159,462    \\
Alipay  & 498,308  & 2,200,291 & 35,179,371 & 6       & 3,327,205    \\
\bottomrule
\end{tabular}
\end{table}

Following RIM~\cite{qin2021retrieval}, we organize the data such that the oldest instances constitute the search pool, the most recent instances comprised the test set, and the instances in between are assigned to the training set. The retrieval-based methods~\cite{qin2021retrieval,qin2020user} retrieve neighboring samples from the search pool. For \secondcategory~\cite{zhou2018deep,zhou2019deep} (e.g., DIEN), the sequential features (e.g., user behavior) are also generated from the search pool. 


\subsubsection{Evaluation Metrics.}

To measure performance, we utilize the commonly adopted metrics AUC and log-loss, which reflect pairwise ranking performance and point-wise likelihood, respectively. A significance test contrasting the two top-performing methods for each metric is undertaken, with results of significance denoted by an asterisk ($*$).

We follow previous work~\cite{zhou2018deep} to introduce Rel. Impr. metric to quantify the relative improvement of models, which is defined as below:
\begin{equation}
    \text { Rel.\ Impr. }=\left(\frac{\mathrm{AUC}(\text {measured model})-0.5}{\text { AUC(base model})-0.5}-1\right) \times 100 \%,
\end{equation}
where the base model is the backbone model of  
 \name in each dataset.

\subsubsection{Compared Baselines.}
For baselines, we compare \name with a mid-tier performance backbone model against traditional models and retrieval-based models. For traditional CTR models, GBDT~\cite{chen1996data} is a frequently used tree-based model, and DeepFM~\cite{guo2017deepfm} is a popular deep-learning model that focuses on feature interactions. HPMN~\cite{ren2019lifelong} and MIMN~\cite{pi2019practice} utilize memory network architectures, while DIN~\cite{zhou2018deep} and DIEN~\cite{zhou2019deep} are attention-based CTR models designed for pinpointing user interests. Additionally, FATE~\cite{wu2021towards} is a model for learning representations of tabular data which facilitates interactions among samples within a minibatch. In item-level retrieval-based models, SIM~\cite{pi2020search} and UBR~\cite{qin2020user} extract pertinent user behaviors from a comprehensive set of user-generated data. 
Regarding the sample-level retrieval-based model, RIM~\cite{qin2021retrieval} retrieves relevant data instances based on raw input features. 
We do not include DERT~\cite{zheng2023dense} in our comparison because it is a special case of RIM that introduces a new dense retrieval setting.
Moreover, our primary goal is to develop a method that eliminates the need for retrieval during inference, thereby improving efficiency and enabling real-world deployment. Therefore, we focus more on comparing the performance of \name with the teacher model \ie RIM~\cite{qin2021retrieval} to determine whether \name would cause a significant performance degradation and whether it could enhance the backbone model.

\subsection{Overall Performance Comparison: RQ1}
\label{performance}
In this section, we compare the overall experimental results presented in Table \ref{tab: overall performance} and make the following observations.
First, \name significantly enhances the performance of the backbone model, with AUC improvements of 10.08\%, 34.96\%, and 16.85\% on the three datasets, respectively, demonstrating its superior performance.
Second, retrieval-based methods~\cite{qin2020user,qin2021retrieval} considerably outperform \secondcategory~\cite{zhou2018deep,zhou2019deep}, benefiting from their advanced knowledge retrieval capabilities. Furthermore, \fourthcategory~\cite{qin2021retrieval} exhibit better performance than \thirdcategory~\cite{pi2020search,qin2020user} due to more comprehensive retrieval strategies.
Finally, \name notably surpasses the performance of \thirdcategory methods, such as SIM~\cite{pi2020search} and UBR~\cite{qin2020user}, and even outperforms our teacher model, RIM~\cite{qin2021retrieval}, which belongs to the \fourthcategory. Our primary goal was to develop a method that eliminates the need for retrieval during inference, thereby improving efficiency and enabling real-world deployment. Initially, we expected that our approach would merely approximate the performance of the teacher model while preserving extraordinary inference efficiency, given that it is based on knowledge distillation. However, through the introduction of contrastive regularization, our method not only matches but exceeds the performance of the teacher model. This demonstrates the strong potential of our approach for real-world applications, as it can effectively learn and utilize the knowledge in the search pool without the need to perform real retrieval during inference. Our results highlight the effectiveness of knowledge distillation combined with contrastive regularization in capturing and leveraging the information from the search pool, enabling efficient and high-performing models suitable for deployment in practical scenarios.

\begin{table*}[]
\centering
\caption{Performance comparison of CTR prediction task with various baselines. For each dataset, RIM is selected as the teacher model, and an average-performing backbone model is selected for \name: DIEN for both the Tmall and Taobao datasets and DIN for the Alipay dataset. $^{\circ}$ indicates recent results. Evaluation metrics include AUC and log-loss (LL). The best results are highlighted in bold, while the runner-ups are underlined. ``Rel. Impr.'' signifies the model's relative AUC improvement over the chosen backbone model, with a statistical significance level of \( p < 0.01 \).}

\label{tab: overall performance}
\begin{tabular}{@{}cc|ccc|ccc|ccc@{}}
\toprule
\multirow{2}{*}{Category} &
  \multirow{2}{*}{Model} &
  \multicolumn{3}{c|}{Tmall} &
  \multicolumn{3}{c|}{Taobao} &
  \multicolumn{3}{c}{Alipay} \\ 
\cmidrule(l){3-11} 
& & AUC & LL & Rel. Impr & AUC & LL & Rel. Impr & AUC & LL & Rel. Impr \\ 
\midrule
\multirow{4}{*}{\begin{tabular}[c]{@{}c@{}}\\\\Traditional\\ Models\end{tabular}} 
 &GBDT & 0.8319 & 0.5103 & -13.55\% & 0.6134 & 0.6797 & -54.75\% & 0.6747 & 0.9062 & -34.00\% \\
 & DeepFM & 0.8581 & 0.4695 & -6.72\% & 0.671 & 0.6497 & -31.76\% & 0.6971 & 0.6271 & -25.54\% \\
& FATE & 0.8553 & 0.4737 & -7.45\% & 0.6762 & 0.6497 & -29.69\% & 0.7356 & 0.6199 & -10.99\% \\
  &HPMN & 0.8526 & 0.4976 & -8.15\% & 0.7599 & 0.5911 & 3.71\% & 0.7681 & 0.5976 & 1.28\% \\
& MIMN & 0.8457 & 0.5008 & -9.95\% & 0.7533 & 0.6002 & 1.08\% & 0.7667 & 0.5998 & 0.76\% \\
& DIN & 0.8796 & 0.4292 & -1.12\% & 0.7433 & 0.6086 & -2.91\% & 0.7647 & 0.6044 & 0.00\% \\
& DIEN & 0.8839 & 0.4272 & 0.00\% & 0.7506 & 0.6082 & 0.00\% & 0.7485 & 0.6019 & -6.12\% \\ 
\midrule
\multirow{3}{*}{\begin{tabular}[c]{@{}c@{}}Retrieval-based\\ Models\end{tabular}} &
  SIM (Item-level)& 0.8857 & 0.4520 & 0.47\% & 0.7825 & 0.5795 & 12.73\% & 0.7600 & 0.6089 & -1.78\% \\
& UBR (Item-level)& 0.8975 & 0.4368 & 3.54\% & 0.8169 & 0.5432 & 26.46\% & 0.7952 & 0.5747 & 11.52\% \\
& RIM (Teacher model)& \underline{0.9151} & \underline{0.3697} & 8.13\% & \textbf{0.8567}$^*$ & \textbf{0.4546}$^*$ & 42.34\% & \underline{0.8005} & \underline{0.5736} & 13.52\% \\
\midrule
\begin{tabular}[c]{@{}c@{}}Our Model\end{tabular} &
  \name & \textbf{0.9226}$^*$ & \textbf{0.3546}$^*$ & 10.08\% & \underline{0.8382} & \underline{0.5098} & 34.96\% & \textbf{0.8093}$^*$ & \textbf{0.5304}$^*$ & 16.85\% \\ 
\bottomrule
\end{tabular}
\end{table*}

\subsection{Compatibility Analysis: RQ2}
\label{sec:Compatibility}
In this section, we begin by contrasting the compatibility of our novel approach, \name, with conventional CTR models, as delineated in Table~\ref{tab:compatibility}. Upon examining the data, two observations arise:
(i) The application of \name markedly amplifies the performance of the backbone models. 
To illustrate, by integrating \name with DeepFM~\cite{guo2017deepfm}, DIN~\cite{zhou2018deep}, and DIEN~\cite{zhou2019deep}, there is a notable enhancement in performance metrics, with an average AUC increase of 4.30\%, 8.01\%, and 5.59\%, coupled with an average decrease in log-loss by 11.80\%, 8.51\%, and 8.22\% across three datasets, respectively. This highlights the significant benefits of incorporating the retrieval-oriented knowledge from \name into the backbone models and indicates that the neural knowledge model successfully serves as a compact surrogate for the search pool.
(ii) \name, exhibiting model-agnosticism, seamlessly integrates with a diverse array of backbone CTR models, irrespective of whether they focus on feature interaction or behavior modeling. This underscores that the knowledge garnered by \name is beneficial across diverse models, highlighting the compatibility of \name. The feature of model-agnosticism is crucial since, regardless of the optimal model for a given scenario, \name consistently boosts its capabilities. Thus, instead of being bound to one model, we can choose the ideal one for each situation and leverage \name to optimize its performance.

\begin{table*}[]
\centering
\caption{Compatibility analysis of applying \name to different backbone models. ``Rel. Impr.'' means relative AUC and log-loss improvements of \name against the backbone. Improvements are statistically significant with \(p< 0.01\).}
\label{tab:compatibility}
\resizebox{\textwidth}{!}{%
\begin{tabular}{@{}c|cccc|cccc|cccc@{}}
\toprule
\multirow{2}{*}{Model} & \multicolumn{4}{c|}{Tmall} & \multicolumn{4}{c|}{Taobao} & \multicolumn{4}{c}{Alipay} \\ 
\cmidrule(l){2-5} \cmidrule(l){6-9} \cmidrule(l){10-13}
                       & AUC & Rel. Impr. & LL & Rel. Impr. & AUC & Rel. Impr. & LL & Rel. Impr. & AUC & Rel. Impr. & LL & Rel. Impr. \\ 
\midrule
DeepFM & 0.8585 & 4.21\% & 0.4803 & 9.31\% & 0.6710 & 3.52\% & 0.6497 & 2.51\% & 0.6971 & 5.59\% & 0.6271 & 4.80\% \\
DeepFM+\name & \textbf{0.8946}$^*$ & - & \textbf{0.4356}$^*$ & - &  \textbf{0.6946}$^*$ & - & \textbf{0.6334}$^*$ & - & \textbf{0.7361}$^*$ & - & \textbf{0.5970}$^*$ & - \\
\midrule
DIN & 0.8796 & 4.26\% & 0.4292 & 5.59\% & 0.7433 & 8.85\% & 0.6086 & 6.82\% & 0.7647 & 5.83\% & 0.6044 & 12.24\% \\
DIN+\name & \textbf{0.9171}$^*$ & - & \textbf{0.4052}$^*$ & - & \textbf{0.8091}$^*$ & - & \textbf{0.5671}$^*$ & - & \textbf{0.8093}$^*$ & - & \textbf{0.5304}$^*$ & - \\
\midrule
DIEN & 0.8839 & 4.38\% & 0.4272 & 20.50\% & 0.7506 & 11.67\% & 0.6082 & 16.18\% & 0.7485 & 5.33\% & 0.6019 & 7.61\% \\
DIEN+\name & \textbf{0.9226}$^*$ & - & \textbf{0.3546}$^*$ & - & \textbf{0.8382}$^*$ & - & \textbf{0.5098}$^*$ & - & \textbf{0.7884}$^*$ & - & \textbf{0.5561}$^*$ & - \\
\bottomrule
\end{tabular}%
}
\end{table*}


\subsection{Comparison of the learned knowledge of \name and RIM: RQ3}
\label{sec:rq3}
In this section, we delve into the comparative analysis of the knowledge quality acquired by \name and RIM.
For a detailed assessment, we formulate two simple variants (LR and MLP) for each model to fully demonstrate the effect of knowledge:
\begin{list}{\labelitemi}{\leftmargin=1em}
\setlength{\topmargin}{0pt}
\setlength{\itemsep}{0em}
\setlength{\parskip}{0pt}
\setlength{\parsep}{0pt}
\item For \name, we design the ROK(LR) and ROK(MLP) variants. Both take as input the concatenation of the original feature embeddings $\mathbf{x}_t$, with \name's knowledge component $\mathbf{z}^x_t$.
\item Similarly, for RIM~\cite{qin2021retrieval}, we introduced the RIM(LR) and RIM(MLP) variants. Their input comprises the concatenation of the original feature embeddings, $\mathbf{x}_t$, with RIM's aggregated features and labels, $\mathbf{r}_t$.
\end{list}
The detailed performance comparison of these variants is tabulated in Table \ref{tab:roklr} and we make two observations.
First, significantly, the results indicate that the \name variants surpass the RIM variants in performance across both datasets. This distinction in outcomes emphasizes the superior quality of knowledge learned by \name, especially as these models apply this intrinsic knowledge directly for predictions. It is compelling to note that while \name leverages the retrieval imitation strategy (detailed in Section \ref{sec:retrieval imitation}) to imitate the aggregated features and labels of RIM, it extracts richer and more profound knowledge from the data. We believe that the contrastive regularization methodology is pivotal in driving this enhanced performance for \name.
Second, ROK(MLP) in Tmall \textbf{even outperforms DeepFM~\cite{guo2017deepfm} and some \secondcategory}, which shows another way to improve the pre-ranking model effect.

\begin{figure*}
    \centering
    \begin{minipage}{0.5\textwidth}
        \begin{subfigure}{\linewidth}
            \includegraphics[width=0.49\linewidth]{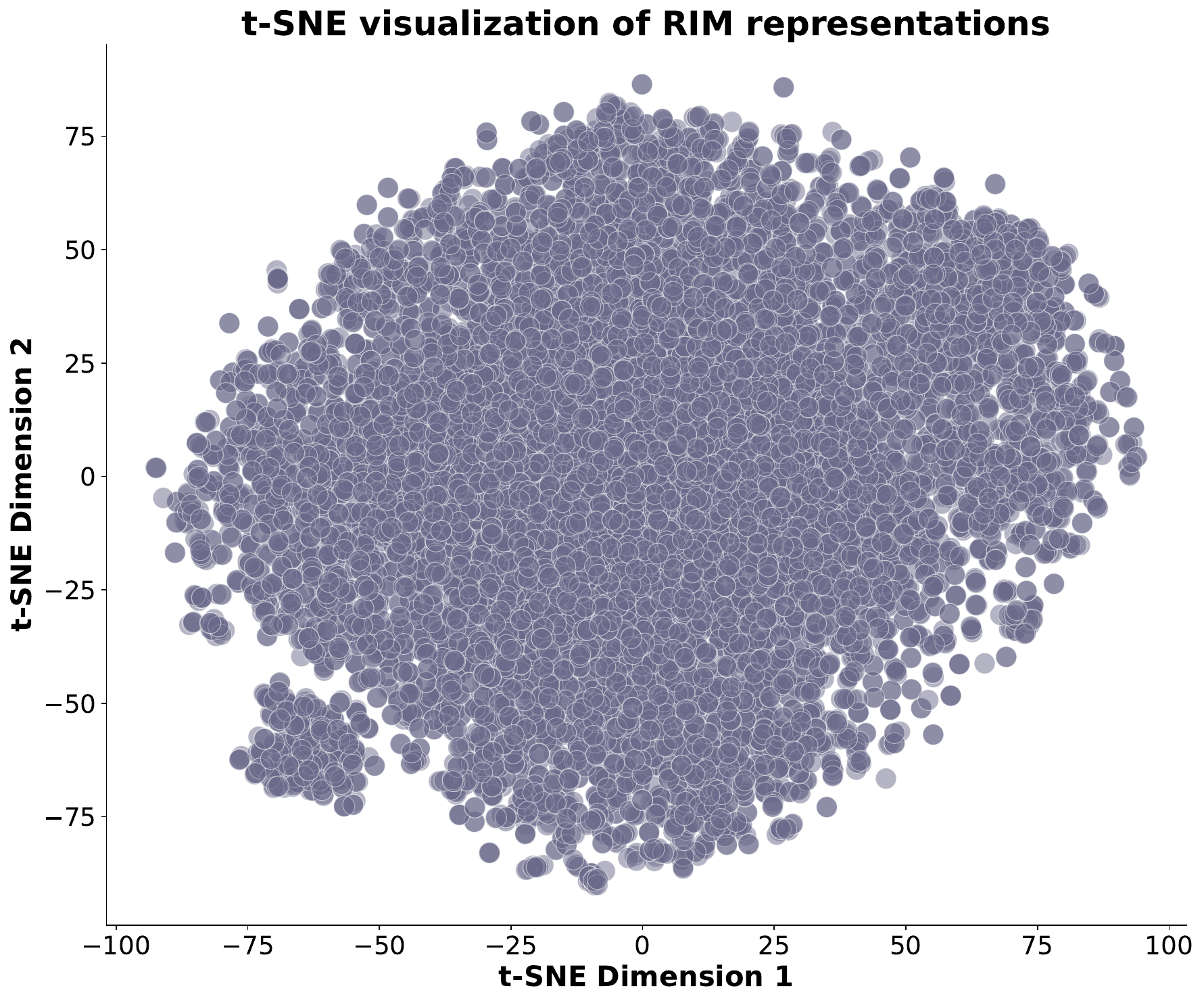}
            \includegraphics[width=0.49\linewidth]{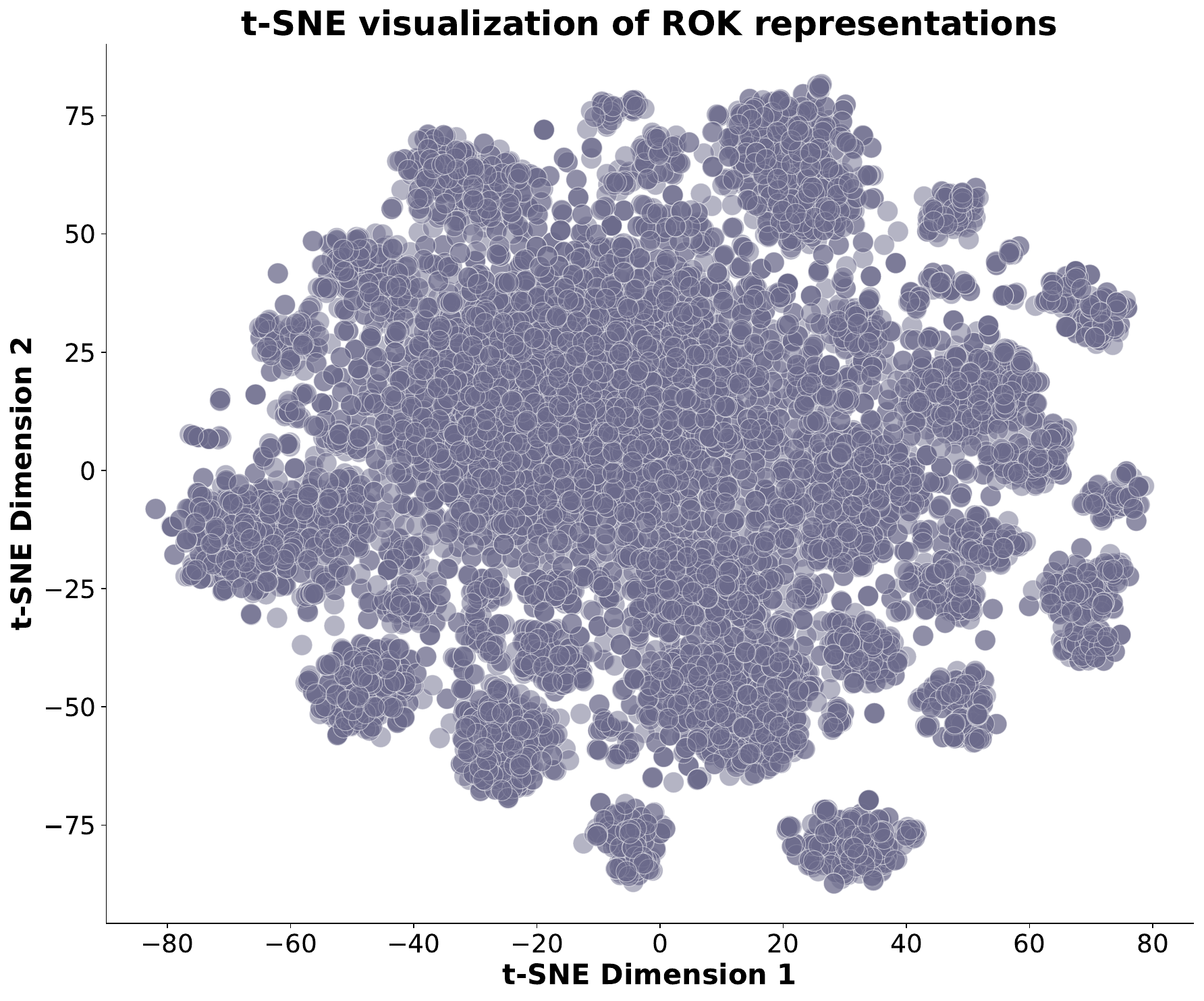}
        \end{subfigure}
        \subcaption{Alipay (Train)}
    \end{minipage}%
    \begin{minipage}{0.5\textwidth}
        \begin{subfigure}{\linewidth}
            \includegraphics[width=0.49\linewidth]{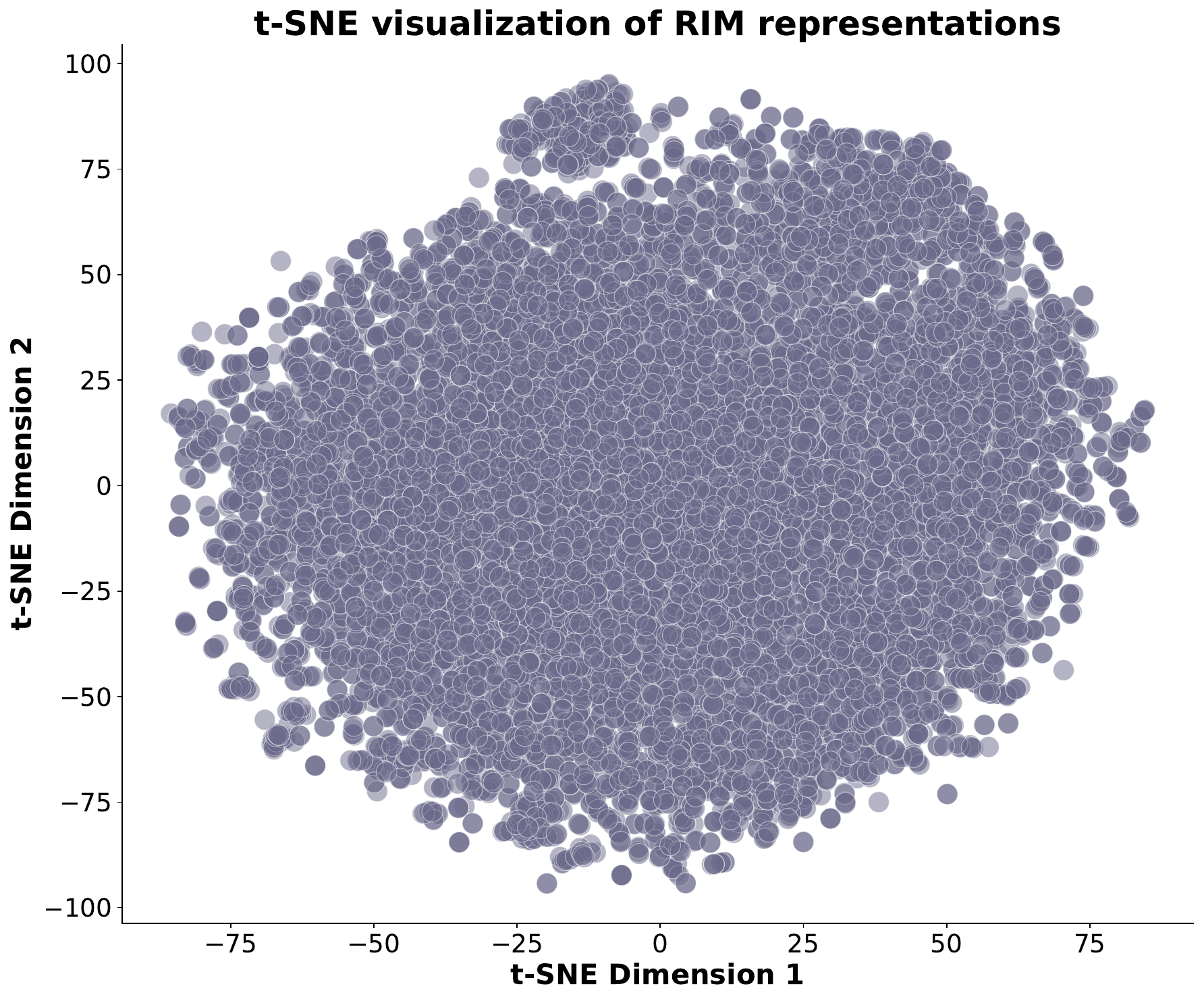}
            \includegraphics[width=0.49\linewidth]{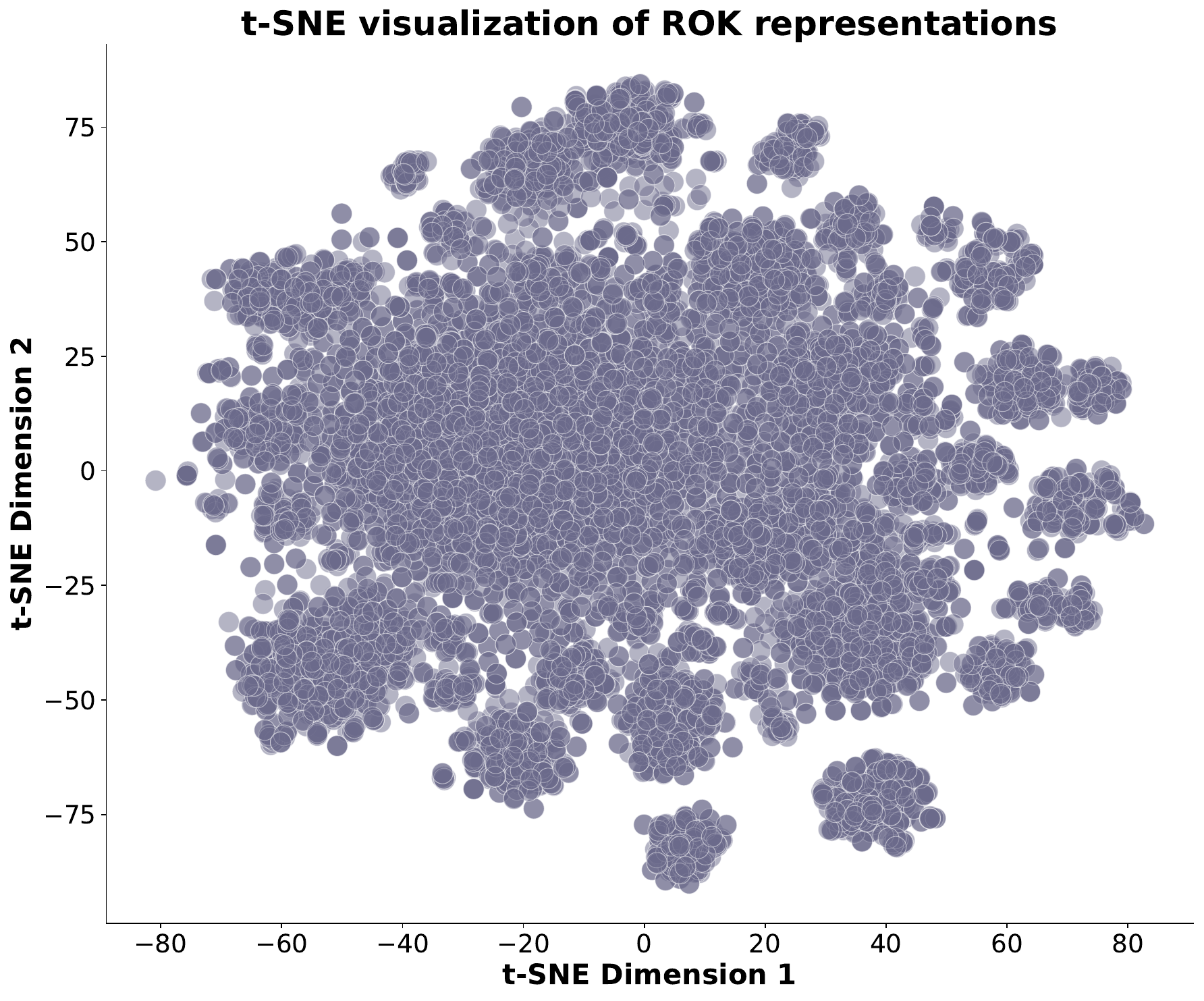}
        \end{subfigure}
        \subcaption{Alipay (Test)}
    \end{minipage}
    
    \caption{The t-SNE visualization of knowledge from RIM (on the left in each subfigure) and \name (on the right in each subfigure) for Alipay.}
    \label{fig:tsne}
\end{figure*}

To delve deeper into the inherent knowledge of \name and explain its superiority over RIM, we employed t-SNE visualization~\cite{van2008visualizing}. Figure~\ref{fig:tsne} illustrates the knowledge distribution patterns of both \name and RIM in the Alipay dataset. For this visualization, we randomly selected $10,000$ samples from both the training and testing sets for each model.
We have two observations. 
First, \name's knowledge distribution exhibits a more pronounced clustering effect compared to RIM. This enhanced clustering can be attributed to the contrastive regularization employed by \name. The contrastive regularization emphasizes the local features of neighboring samples and mitigates the noise by selectively considering the most relevant sample, as shown in section~\ref{sec:Contrastive Regularization}. Specifically, representations of analogous samples are pulled closer, while those of dissimilar samples are pushed apart, leading to a more distinct and meaningful clustering. In contrast, RIM's retrieved neighboring samples may include some incidental noise that correlates with the quantity of these samples, and its aggregation mechanism only captures the global features of the neighboring samples.
Second, the visualization of \name's representations reveals a high degree of similarity between the training and test datasets, demonstrating the good generalization ability of the decomposition-reconstruction paradigm. This is significant because a major concern when using neural networks as a knowledge base is their generalization ability, as the training process is guided by the retrieved examples of \fourthcategory, while the testing phase requires the model to apply its knowledge independently.

\begin{table}[h]
\centering
\caption{Comparison of model variants. ``Rel. Impr.'' means relative AUC improvement of ROK's variants against baselines and RIM's variants. Improvements are statistically significant with $p< 0.01$.}
\label{tab:roklr}
\begin{tabular}{lccccc}
\toprule
Model & \multicolumn{2}{c}{Tmall} & \multicolumn{2}{c}{Alipay} \\
\cmidrule(lr){2-3} \cmidrule(lr){4-5}
 & AUC & Rel. Impr. & AUC & Rel. Impr. \\
\midrule
LR & 0.8213 & 4.37\% & 0.6298 & 8.96\%\\
RIM(LR) & 0.8379 & 2.30\% & 0.6613 & 3.77\%\\
ROK(LR) & \textbf{0.8572}$^*$ & - & \textbf{0.6862}$^*$ & - \\
\midrule
MLP & 0.8393 & 2.98\% & 0.6344 & 9.13\%\\
RIM(MLP) & 0.8426 & 2.58\% & 0.6773 & 2.21\%\\
ROK(MLP) & \textbf{0.8643}$^*$ & - & \textbf{0.6923}$^*$ & - \\
\bottomrule
\end{tabular}
\end{table}


\subsection{Ablation \& Hyper-parameter Study}
\vspace{-5pt}
\subsubsection{Update Strategy: RQ4.}
\label{sec: update_strategy}
\begin{figure*} 
    \centering
    \begin{subfigure}{0.3\linewidth} 
        \centering
        \includegraphics[width=\linewidth]{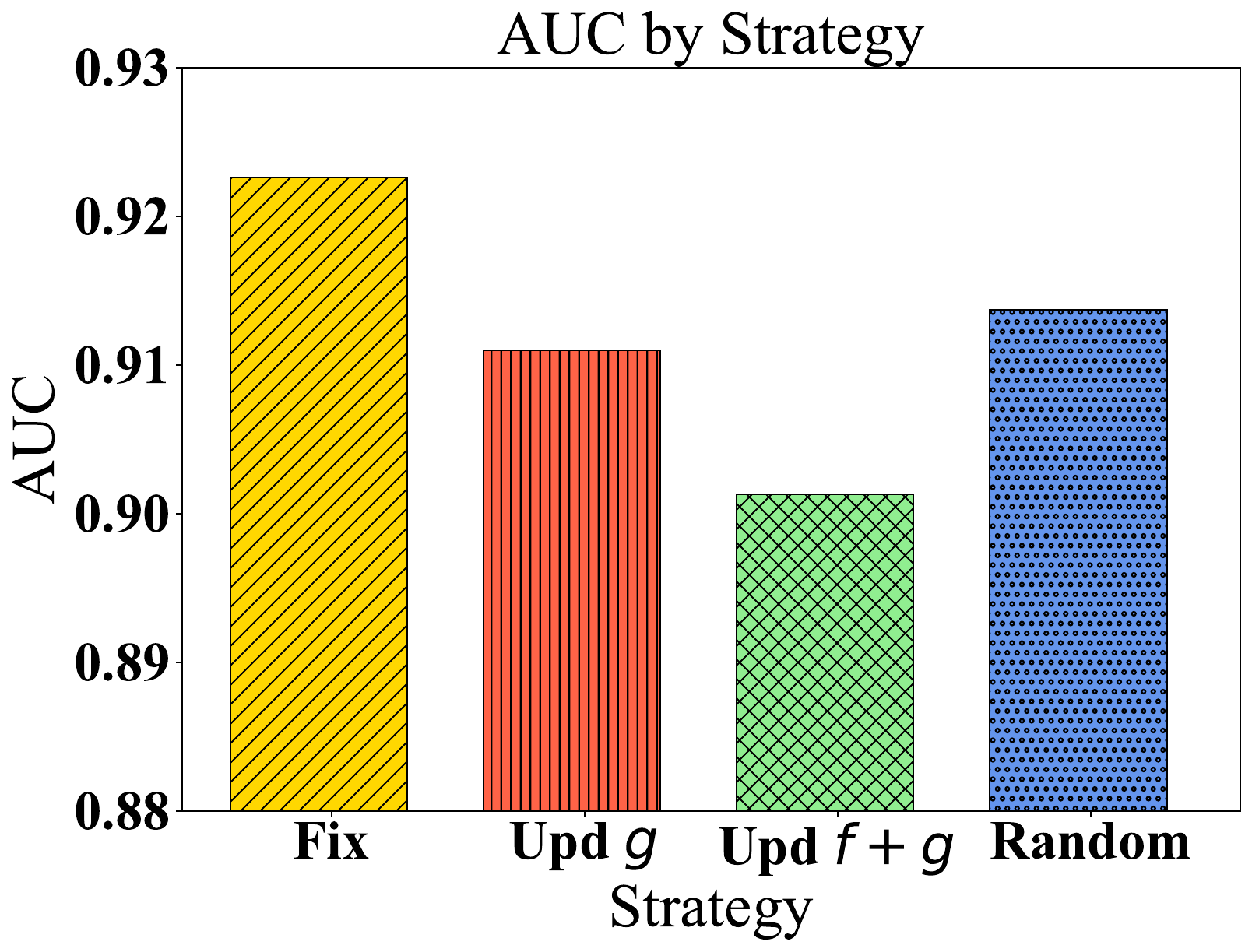} 
        \caption{Tmall} 
    \end{subfigure}
    \hfill
    \begin{subfigure}{0.3\linewidth} 
        \centering  
        \includegraphics[width=\linewidth]{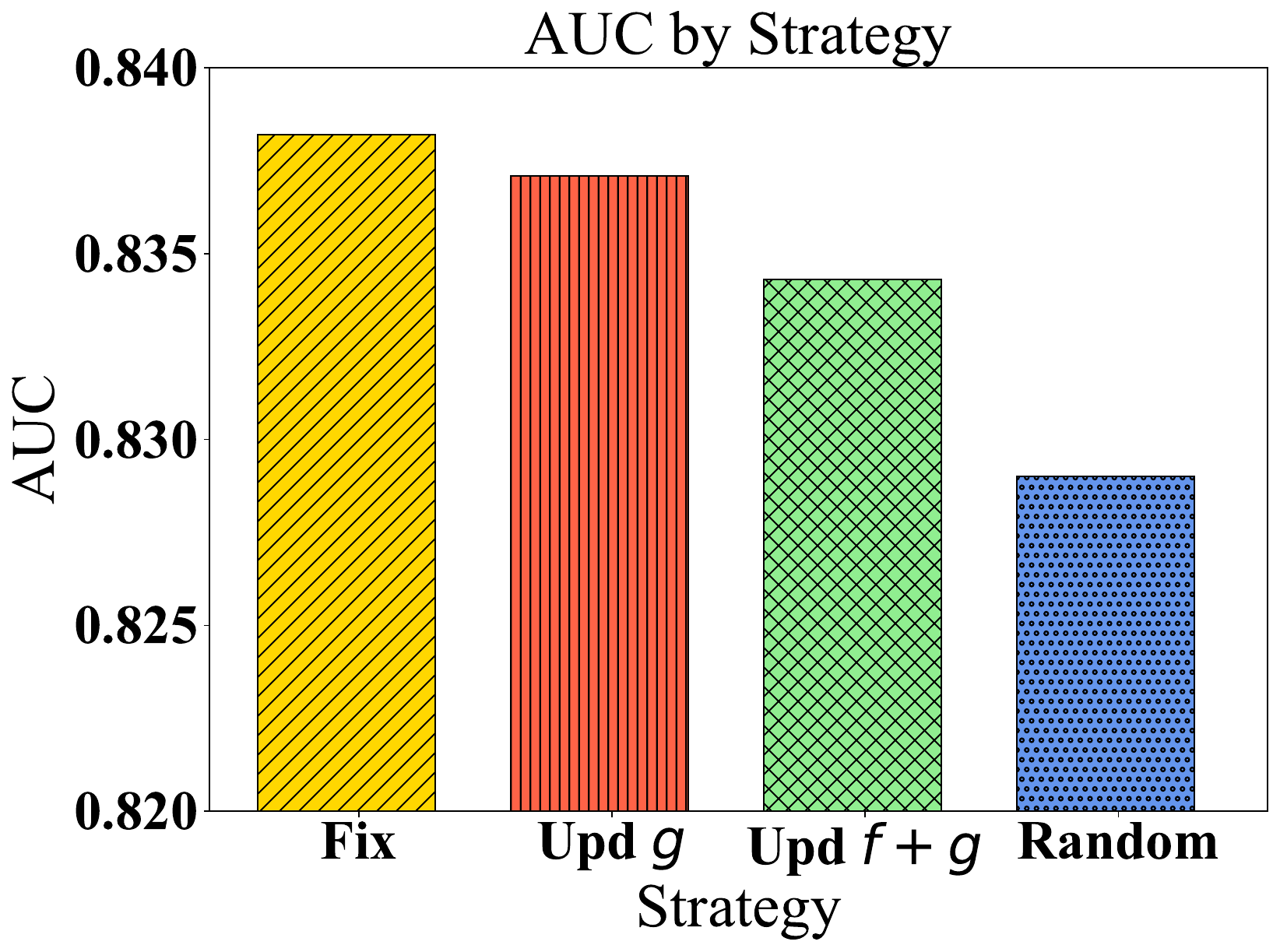} 
        \caption{Taobao} 
    \end{subfigure}
    \hfill
    \begin{subfigure}{0.3\linewidth} 
        \centering
        \includegraphics[width=\linewidth]{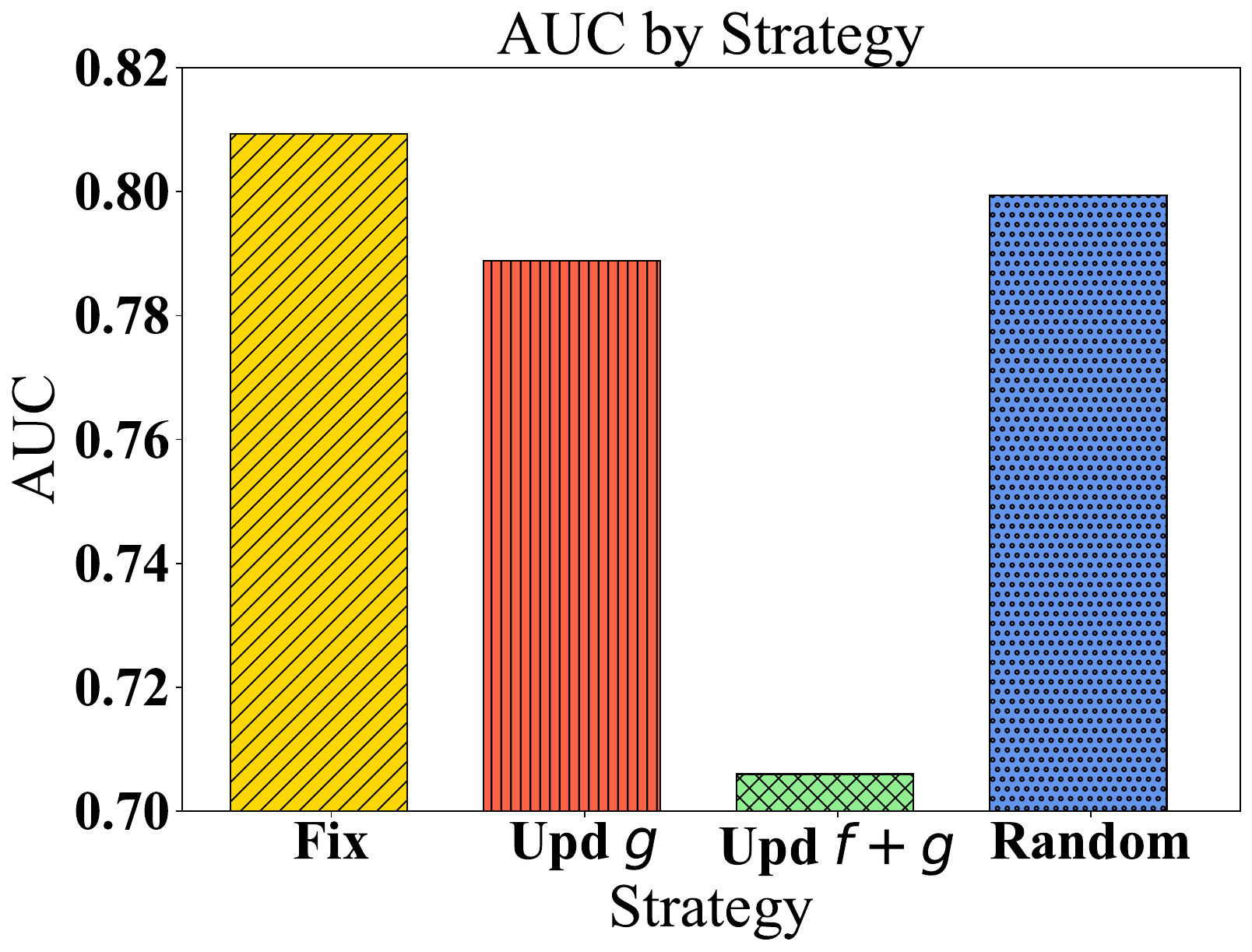} 
        \caption{Alipay} 
    \end{subfigure}
    \caption{Comparison of AUC scores for different update strategies.} 
     \label{fig: update}
\end{figure*}

In this section, we undertake comprehensive ablation experiments for \name. We begin by examining the parameter update strategy employed for the knowledge base during the knowledge utilization phase. The strategies under consideration include: fixing the knowledge base (\textit{Fix}), updating only the upper knowledge encoder $g$ (\textit{Upd $g$}), and updating both the knowledge encoder and the retrieval-oriented embedding layer $f$ (\textit{Upd $f+g$}). Furthermore, we consider the positive sample selection strategy described in Section \ref{sec:Contrastive Regularization}. Specifically, we select the most related neighboring sample by default and compare its performance to randomly selecting a sample from the retrieved samples (\textit{random}).
Figure \ref{fig: update} depicts the results obtained on the Tmall, Taobao and Alipay datasets. We have two observations: First, randomly selecting a sample as the positive sample diminishes performance, suggesting that even within the small retrieved set of samples, noise is still present. Second, a key observation is that the strategy of fixing the knowledge base (\textit{Fix}) yields the best performance. This can be attributed to the primary goal of \name: to derive superior representations. However, updating with the backbone models may adversely impact the acquired knowledge.
This characteristic is highly desirable and aligns well with our expectations, further enhancing the practicality of \name. It indicates that once \name has been fully trained, it can function as an independent module, eliminating the need for subsequent modifications and reinforcing its role as an authentic knowledge base.



\subsubsection{Hyperparameter Study: RQ5.}
\label{sec:hyperparameter study}
\begin{figure} 
    \centering
    \begin{subfigure}{0.49\linewidth} 
        \centering
        \includegraphics[width=\linewidth]{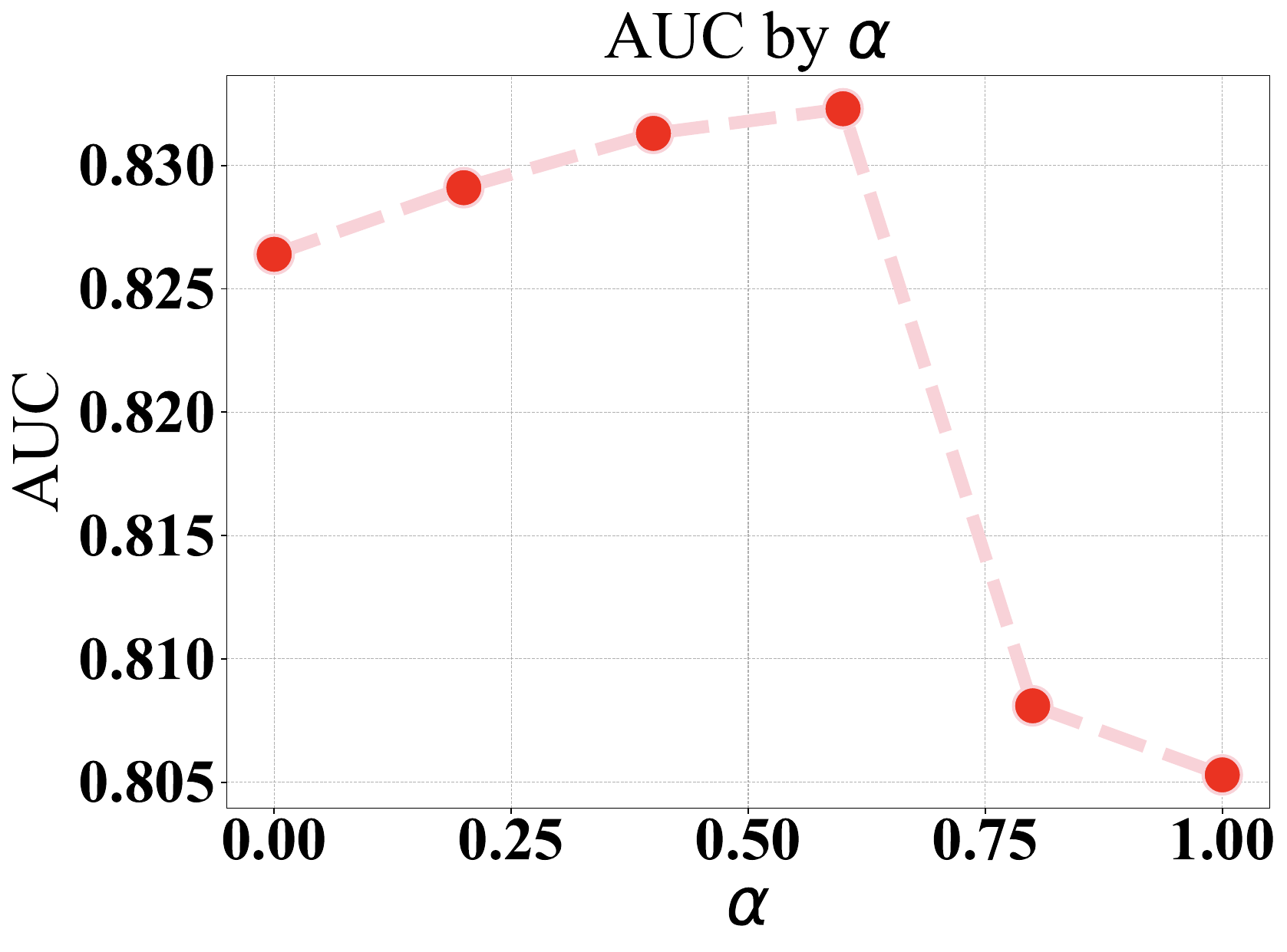} 
        \caption{Taobao} 
    \end{subfigure}
    \hfill
    \begin{subfigure}{0.49\linewidth} 
        \centering
        \includegraphics[width=\linewidth]{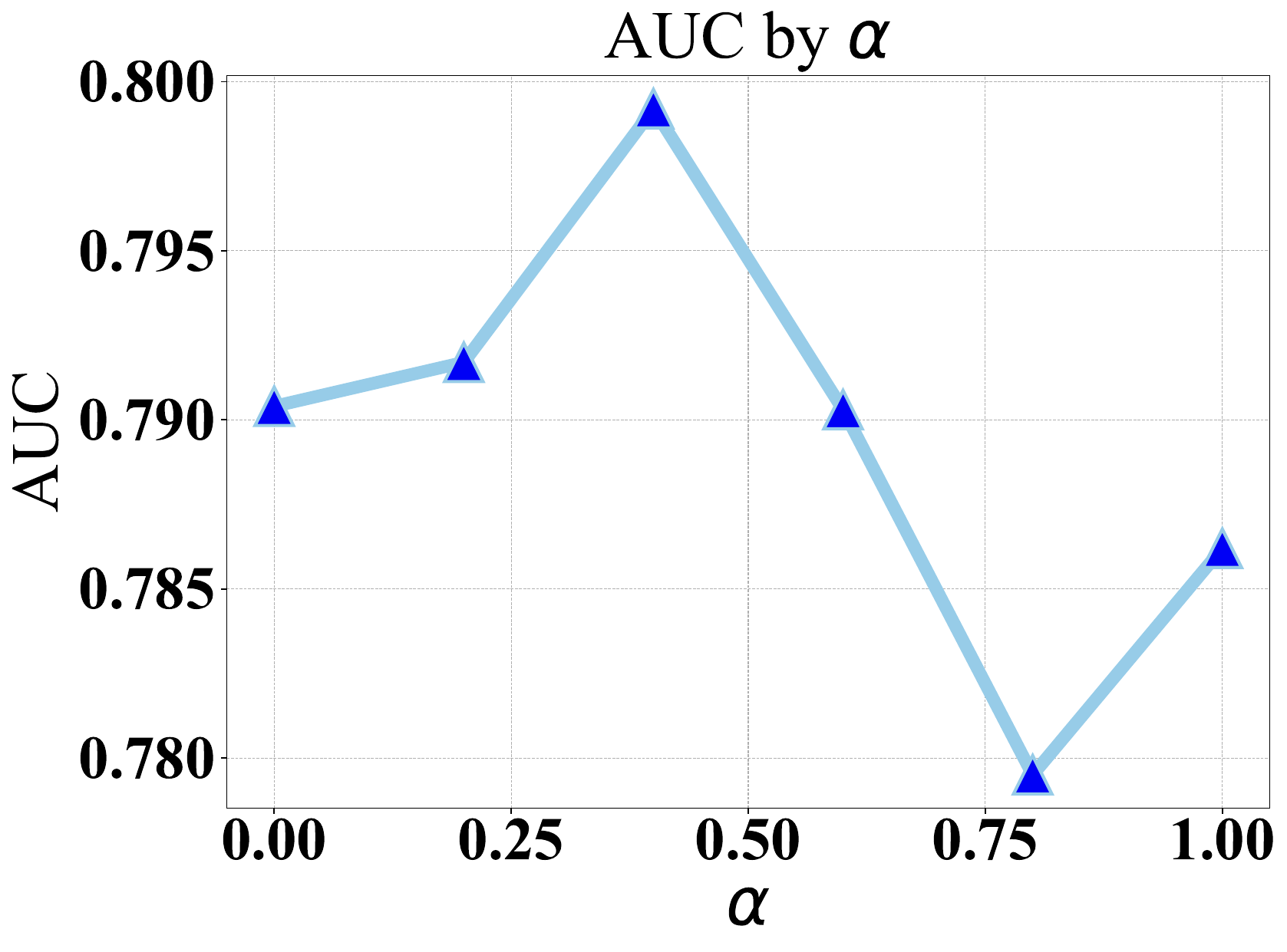} 
        \caption{Alipay} 
    \end{subfigure}
    \caption{Comparison of AUC scores across varying $\alpha$ values.} 
    \label{fig: alpha}
\end{figure}

In this section, we delve into the study of hyperparameter $\alpha$ pertaining to the balance between the contrastive regularization loss and retrieval imitation loss, represented as \( \alpha \).
As illustrated in Figure~\ref{fig: alpha}, for both datasets, when \( \alpha \) equals either \( 0 \) or \( 1 \)—representing the extreme scenarios—the performance of \name deteriorates. This observation underscores the significance of both retrieval imitation loss and contrastive regularization loss to the model's efficacy.

%% file: 4Deployment.tex
\section{Industry Application}
\subsection{Training Efficiency}
Training \name involves pre-training a sample-level retrieval-based model, creating retrieval-oriented knowledge, and applying this knowledge to backbone CTR models. Despite these steps, the extra training time is reasonable compared to training a standalone model. The training duration is detailed in Table~\ref{tab:training_time}. 
Two observations stand out. Firstly, incorporating \name into the backbone model significantly reduces training time, thanks to fewer parameters needing updates, partly because \name's parameters are frozen and the embedding size is halved. This leads to faster convergence. 
Second, the total time of Phase 1 and Phase 2 is close to the training duration of the DIEN without \name. The relatively time-consuming part of the whole training process would be the RIM's retrieval mechanism. However, this is mitigated by using large bulk sizes and caching, which decreases retrieval times. For further efficiency, extending the knowledge base update frequency beyond that of the backbone model could be beneficial.

\begin{table}[H]
\centering
\caption{Training duration comparison for two phases, total time and backbone model on Tmall, Taobao, and Alipay, measured in minutes. Phase 1 includes data retrieval for training and testing tests, pre-training RIM, and Retrieval-Oriented Knowledge Construction. Phase 2 includes Knowledge Utilization, with DIEN as the backbone model. Additionally, DIEN training times without ROK integration are provided. For Phase 1, data retrieval times are specified separately.}
\label{tab:training_time}
\begin{tabular}{lccc}
\hline
Phase & Tmall & Taobao & Alipay \\ \hline
Phase 1 (Retrieval)  & 156 (40)          & 116 (30)           & 24 (4)            \\
Phase 2 & 201      & 163         & 68           \\ 
\midrule
Total & 357      & 279         & 92\\
DIEN without \name & 327 & 330 & 79\\ \hline

\end{tabular}
\end{table}
\vspace{-10pt}

\subsection{Inference Efficiency} \label{ssec:Inference_Efficiency}

Deploying \fourthcategory is complex and costly due to several key factors like extreme inefficiency problems during inference, high resource consumption and divergent inference process that requires an inference-retrieval-inference cycle.
Our method, \name, introduces a knowledge base that improves retrieval efficiency, reducing time complexity to $\mathcal{O}(1)$. We assessed backbone models—DeepFM, DIEN, and DIN—integrated with \name and compared them to UBR~\cite{qin2020user}, sparse \fourthcategory, and dense \fourthcategory. As shown in Table~\ref{tab:testspeed}, retrieval-based methods have significantly longer inference times, making them impractical for real-time requirements.
In contrast, \name improves backbone models with minimal additional inference time, making it suitable for online deployment. Table~\ref{tab:compatibility} and Table~\ref{tab:testspeed} demonstrate that \name consistently improves the AUC of all backbone models across the tested datasets while only slightly increasing the inference time by 0.12 ms to 0.25 ms per sample, showcasing its ability to enhance performance without significantly impacting speed.

\name's retrieval-free design ensures compatibility with existing systems, requiring only one inference process, thus avoiding data transfer overhead and enhancing efficiency. This combination of accuracy and rapid inference makes \name a superior solution for modern online services.

\begin{table}[h]
\centering
\caption{Comparison of inference speed (ms per sample) across models on Tmall, Taobao, and Alipay. For UBR, sparse retrieval and dense retrieval, both total inference and retrieval times are presented to highlight the impact of the retrieval process.}
\label{tab:testspeed}
\resizebox{\linewidth}{!}{
\begin{tabular}{lccc}
\toprule
\multirow{1}{*}{Model} & \multicolumn{1}{c}{Tmall} & \multicolumn{1}{c}{Taobao} & \multicolumn{1}{c}{Alipay} \\
\midrule
UBR (Retrieval) & 20.71 (17.67) & 56.45 (53.32) & 30.32 (27.31) \\
Sparse (Retrieval) & 174.81 (173.27) & 206.22 (204.78) & 113.95 (112.37) \\
Dense (Retrieval) & 17.78 (16.19) & 19.53 (16.97) & 17.95 (16.31) \\
\midrule
DeepFM & 1.34 & 1.36 & 1.19 \\
DeepFM+\name & 1.46 & 1.40 & 1.41 \\
\midrule
DIEN & 5.44 & 4.89 & 3.69 \\
DIEN+\name & 5.51 & 5.04 & 3.73 \\
\midrule
DIN & 1.28 & 1.32 & 1.43 \\
DIN+\name & 1.46 & 1.36 & 1.56 \\
\bottomrule
\end{tabular}
}
\end{table}

%% file: 5Conslusion.tex
\section{Conclusion}


The \name framework tackles the critical challenges faced by sample-level retrieval-based methods, such as inference inefficiency and high resource consumption. By introducing a neural network-based Knowledge Base, \name effectively distills and stores retrieval-oriented knowledge from pre-trained retrieval-based models. This innovative approach enables the seamless integration of retrieval-enhanced representations with various CTR models at both the instance and feature levels, bypassing the need for time-consuming retrieval during inference.
Extensive experiments on three large-scale datasets demonstrate \name's exceptional compatibility and performance, significantly enhancing the performance of various CTR methods. Remarkably, the introduction of contrastive regularization allows \name to surpass the performance of the sample-level retrieval-based teacher model, highlighting its strong potential for real-world applications.
The \name framework successfully transforms sample-level retrieval-based methods into practical, efficient, and scalable solutions for industrial deployment. By eliminating the need for resource-intensive retrieval during inference and demonstrating the feasibility of distilling knowledge from non-parametric models using a parametric approach, \name opens up new possibilities for leveraging vast amounts of historical data in CTR prediction.

\section{Acknowledgement}
We thank MindSpore~\citep{mindspore} for the partial support of this work, which is a new deep learning computing framework.

